\documentclass[a4paper,11pt]{article}
\usepackage{pos}
\usepackage{subfigure}
\title{Noisy SUSY}

\author*{Stam Nicolis}

\affiliation{Institut Denis Poisson,\\
  Université de Tours, Université d'Orléans, CNRS(UMR7013),\\ Parc Grandmont, 37200 Tours, France}

\emailAdd{Stam;Nicolis@lmpt.univ-tours.fr}

\abstract{
We review the idea, put forward in 1982, by Parisi and Sourlas, that the bath of fluctuations, with which a physical system is in equilibrium, can be resolved by the superpartners 
of the degrees of freedom, defined by the classical action. This implies, in particular, that fermions can be described in terms of their superpartners, using the Nicolai map. 
We focus on the question, whether the fluctuations of scalar fields can, in fact, produce the absolute value of the stochastic determinant itself, whose contribution to the action can be identified with the fermionic degrees of freedom and present evidence supporting this idea in two spacetime dimensions. The same idea leads to a new formulation of supersymmetric QED.
We also review the obstacles for extending this approach to Yang-Mills theories and report on progress for evading the obstructions for obtaining interacting theories in three and four spacetime dimensions. 
This implies, in particular, that it is possible to describe the effects of fermions in numerical simulations, through their superpartners.
}

\FullConference{%
  Corfu Summer Institute 2022 "School and Workshops on Elementary Particle Physics and Gravity",\\
  28 August - 1 October, 2022\\
  Corfu, Greece}

\begin{document}
\maketitle

\section{Introduction}\label{intro}
2022 is a jubilee year: For the Corfu Workshops, that started in 1982; and for the paper ``Supersymmetric Field Theories and Stochastic Differential Equations'' by G. Parisi and N. Sourlas~\cite{parisi_sourlas} that appeared and was published that year. 

In that paper the authors proposed that supersymmetry isn't an optional feature of Nature but that it  expresses a fundamental property of any physical system, in equilibrium with a bath of fluctuations in the sense that the superpartners of the degrees of freedom, that define the physical system, resolve the bath of fluctuations.  

There are several issues that were left unresolved in their presentation: 
\begin{itemize}
\item Can the fluctuations of physical systems be, indeed, ``repackaged'' in the absolute value of the determinant, that defines the change of variables in the path integral, from the ``dynamical'' fields, that describe the physical system, in the absence of fluctuations, to the ``noise'' fields, that describe the bath? Or must the dynamical fields that resolve the fluctuations, be introduced by hand? This is, in fact, what has been used by the studies on the lattice of supersymmetric theories, that do use the Nicolai map. We would like to present evidence that it's possible to simplify the calculations considerably. This is what, indeed, Parisi and Sourlas stressed in their paper.
\item The approach by Parisi and Sourlas  seemed to hit an obstruction beyond two spacetime dimensions; however it hasn't been fully clarified since, whether this is an obstacle of principle for the approach itself, or a technical issue. 
\item The approach was for Wess--Zumino models and left open the question of how to describe gauge theories. 
\item For the case of target space supersymmetry, the Nicolai map--or the Langevin equation--(they are the same relation)  takes the following form (we work throughout in Euclidian signature, in $d$ spacetime dimensions)
\begin{equation}
\label{langevin-nicolai}
\eta_I=\sigma_{IJ}^\mu\partial_\mu\phi_J + \frac{\partial W}{\partial\phi_I}
\end{equation}
where the $\sigma^\mu$ are assumed to generate a Clifford algebra, 
\begin{equation}
\label{Clifford}
\left\{\sigma^\mu,\sigma^\nu\right\}_{IJ}=2\delta^{\mu\nu}\delta_{IJ}.
\end{equation}
Now the generators of a Clifford algebra carry {\em spinor} indices, which would mean that $\eta_I(x)$ and $\phi_I(x)$ are components of spinors; however we would really like to identify the $I,J$ indices as flavor indices. On the other hand, we can do so, but then the transformation properties of the noise fields under spacetime rotations (in Euclidian signature) do not have an obvious interpretation. So there is a conceptual issue here that requires clarification. It is possible to take a ``pragmatic'' approach, whereby we take the $\sigma_{IJ}^\mu$ to be numerical coefficients and try to check the consistency of this map by the identities that the $\eta_I(x),$ as functions of the $\phi_I(x),$ are expected to satisfy; if these don't show any anomalies, this is an indication that the approach is, in fact, consistent. This is what we shall do here. We shall provide evidence that the ``pragmatic'' approach does seem to work. 

Let us, also, note that, in Euclidian signature, the $\sigma_{IJ}^\mu,$ that satisfy eq.~(\ref{Clifford}), have real entries iff $d\equiv\,2\,\mathrm{mod}\,8;$ so, for other spacetime dimensions, the Nicolai map is incomplete, since the RHS can take complex values, whereas the LHS can take real values. One solution is to double the degrees of freedom, i.e. to consider complex scalars, $\phi_I(x),\phi_I^\dagger(x)$ and complex noise fields, $\eta_I(x)$ and $\eta_I^\dagger(x),$ with the latter defined by  
\begin{equation}
\label{doublers}
\eta_I^\dagger(x)=\sigma_{JI}^\mu\partial_\mu\phi_J^\dagger(x)+\left(\frac{\partial W}{\partial\phi_I}\right)^\dagger
\end{equation}
since $[\sigma_{IJ}^\mu]^\dagger=\sigma_{JI}^\mu,$ as they're Hermitian matrices.

Indeed, the idea that we would like to propose here is that, while it may seem that  the $\eta_I(x)$ don't have well-defined properties, under spacetime rotations, the combinations, $\eta_I(x)\eta_J(x)\delta^{IJ},$ (when $d\equiv\,2\,\mathrm{mod}\,8;$ the $\eta_I(x)\eta_J^\dagger(x)\delta^{IJ}$ otherwise) are certainly expected to have: They are supposed to be target space  scalars, also upon expressing them in terms of the fields $\phi_I(x)$ (respectively $\phi_I^\dagger(x)$). It is this requirement that will impose upon the superpotential, $W,$ that it be a holomorphic function of its arguments: This is due to the requirement that the action be invariant under spacetime rotations (in Euclidian signature).
\end{itemize}
In this presentation we would like to report on progress on the understanding of these issues, namely we shall provide the calculational framework and some of its applications that will allow us to understand better what these issues entail.

\section{Repackaging the fluctuations in two dimensions}\label{repack} 
Let us recall that the starting point is the partition function for the ``noise'' fields, $\eta(x):$
\begin{equation}
\label{noisefieldsZ}
Z=\int\,[\mathscr{D}\eta(x)]\,e^{-\int\,d^dx\,\frac{\eta_I(x)\eta_J(x)}{2\sigma^2}\delta^{IJ}}=1
\end{equation}
which we can take equal to 1 by a definition of the measure. Here $\sigma^2$ defines the scale of the fluctuations: It can be identified with $k_\mathrm{B}T$ for thermal fluctuations, with $\hbar$ for quantum fluctuations and with the strength of the disorder for the fluctuations described by disorder. Since the bath is fixed, we can choose our units so that $\sigma^2=1,$ which we shall do in the following.

Let us now perform the change of variables defined by eq.~(\ref{langevin-nicolai}). A change of variables doesn't change the value of the integral (in the absence of anomalies, that, in the present context, would be from boundaries), therefore we obtain the expression for the partition function of the fields $\phi_I(x):$
\begin{equation}
\label{phifieldsZ}
Z=\int\,[\mathscr{D}\phi_I]\,\left|\mathrm{det}\frac{\delta\eta_J}{\delta\phi_I}\right|\,e^{-\frac{1}{2}\int\,d^dx\,\left(\sigma_{IK}^\mu\partial_\mu\phi_K + \frac{\partial W}{\partial\phi_I}\right)\left(\sigma_{JL}^\mu\partial_\mu\phi_L + \frac{\partial W}{\partial\phi_J}\right)\delta^{IJ} }=1
\end{equation}
Indeed we will be able to test, whether anomalies can occur, or not.

What this expression implies is that the fluctuations to the classical action are repackaged into the absolute value of the determinant of the operator $\delta\eta_J/\delta\phi_I:$
It is the presence of this term that ensures that $Z=1.$ 

Now the question arises, whether we must include this Jacobian by hand, or whether it can be produced by the fluctuations of the classical action of the scalars, in equilibrium with them. 

What hasn't been appreciated, even though it's the whole point of  the Nicolai map and of the proposal of Parisi and Sourlas, is that the answer to this question can be obtained by checking the identities that are expected to be satisfied by the noise fields, $\eta_I(x),$ defined in terms of the scalars, $\phi_I(x)$ and, therefore, sampled using the action of the scalars alone, that appears in the exponent of eq.~(\ref{phifieldsZ}). 

The Jacobian (along with its absolute value) will be produced by the fluctuations of the classical action of the scalars, iff the noise fields, as functions of the scalars, satisfy the identities 
\begin{equation}
\label{wicketa}
\begin{array}{l}
\langle(\eta_I(x)-\langle\eta_I(x)\rangle)(\eta_J(x')-\langle\eta_J(x')\rangle)\rangle=\delta_{IJ}\delta(x-x')\\
\langle(\eta_{I_1}(x_1)-\langle\eta_{I_1}(x_1)\rangle)\cdots
(\eta_{I_{2n}}(x_{2n})-\langle\eta_{I_{2n}}(x_{2n})\rangle)
\rangle=\\
\sum_{P}\,\langle(\eta_{I_P(1)}(x_{P(1)})-\langle\eta_{I_{P(1)}}(x_{P(1)}))\rangle\cdots\langle(\eta_{I_{P(2n)}}(x_{P(2n)})-\langle\eta_{I_{P(2n)}}(x_{P(2n)}))\rangle
\end{array}
\end{equation}
Notice that, in these expressions the 1-point function(s) of the noise fields, $\langle\eta_I(x_I)\rangle$--expressed as a function(s) of the scalars--need not vanish. If they do, supersymmetry is realized in the Wigner mode; if they don't it's realized in the Nambu-Goldstone mode; if the $\eta_I(x)$ don't satisfy these identities, supersymmetry is anomalously broken. 

What is important is that the expectation values are to be taken with respect to the measure,
\begin{equation}
\label{measure}
\rho(\phi)\equiv\frac{ e^{-\frac{1}{2}\int\,d^dx\,\left(\sigma_{IK}^\mu\partial_\mu\phi_K + \frac{\partial W}{\partial\phi_I}\right)\left(\sigma_{JL}^\mu\partial_\mu\phi_L 
+ \frac{\partial W}{\partial\phi_J}\right)\delta^{IJ} } }
{ \int\,[\mathscr{D}\phi_I]\,e^{-\frac{1}{2}\int\,d^dx\,\left(\sigma_{IK}^\mu\partial_\mu\phi_K + \frac{\partial W}{\partial\phi_I}\right)\left(\sigma_{JL}^\mu\partial_\mu\phi_L + \frac{\partial W}{\partial\phi_J}\right)\delta^{IJ} } }
\end{equation}
that is  perfectly well-defined, in $D\equiv\,2\,\mathrm{mod}\,8$ and, correspondingly, upon doubling the degrees of freedom, for other values of $D,$ therefore is well-suited for numerical simulations. 

This statement, however, relies on the transformation properties of the $\eta_I(x),$ as functions of the $\phi_I(x).$  If the $\eta_I(x)$ are flavor vectors, they certainly don't look like them, when expressed in terms of the $\phi_I(x).$ To understand what the Nicolai map entails, let's expand the action of the scalars: 
\begin{equation}
\label{Sscalars}
\begin{array}{l}
\displaystyle
S[\phi]=\int\,d^dx\,\frac{1}{2}\left\{ \sigma_{IK}^\mu\sigma_{JL}^\nu\partial_\mu\phi_K\partial_\nu\phi_L+\frac{\partial W}{\partial\phi_I}\frac{\partial W}{\partial\phi_J}+
\sigma_{IK}^\mu\partial_\mu\phi_K \frac{\partial W}{\partial\phi_J} +  \frac{\partial W}{\partial\phi_I}\sigma_{JL}^\mu\partial_\mu\phi_L 
\right\}\delta^{IJ}
\end{array}
\end{equation}
The first term is, in fact, invariant under SO(2) rotations of the coordinates and SO(2) flavor transformations. The second term is invariant under SO(2) rotations of the coordinates, since it doesn't involve any derivatives; what  is not obvious is, whether it is invariant under SO(2) flavor transformations. For $d=2$ it will be, provided that $W(\phi_I)$ is given by the following expression:
\begin{equation}
\label{Wd=2}
W(\phi_1,\phi_2)=\frac{m}{2}(\phi_1^2+\phi_2^2)+w(\phi_1+\mathrm{i}\phi_2)+w(\phi_1-\mathrm{i}\phi_2)
\end{equation} 
The reason is that this  expression is a solution to the equation
\begin{equation}
\label{SO2invV}
\left(\phi_1\frac{\partial}{\partial\phi_2}-\phi_2\frac{\partial}{\partial\phi_1}\right)\left\{\frac{\partial W}{\partial\phi_I}\frac{\partial W}{\partial\phi_J}\delta_{IJ}\right\}=0+\mathrm{total\,derivatives}
\end{equation}
What is interesting is that the expression~(\ref{Wd=2}) for the superpotential ensures that the third term, also, is a total derivative, thereby ensuring that the classical action is, indeed, invariant under SO(2) rotations of the coordinates and SO(2) flavor transformations. However, only the first or the second and third terms can appear, not all three.

Therefore the non--trivial statement is that, iff the superpotential is a holomorphic function of the scalars, the action is invariant under SO(2) coordinate rotations and SO(2) flavor rotations. Furthermore, these symmetries can be expressed as the property that the noise fields, $\eta_I(x),$ upon being expressed as function of the scalars, satify the identities~(\ref{wicketa}), computed using the functions of the scalars alone, which, in this case, is bounded from below and confines at infinity.  
 
\begin{figure}[thp]
\subfigure{\includegraphics[scale=0.35]{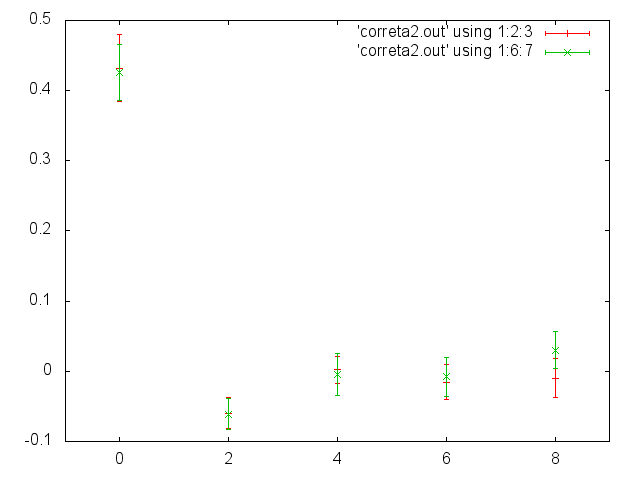}}
\subfigure{\includegraphics[scale=0.35]{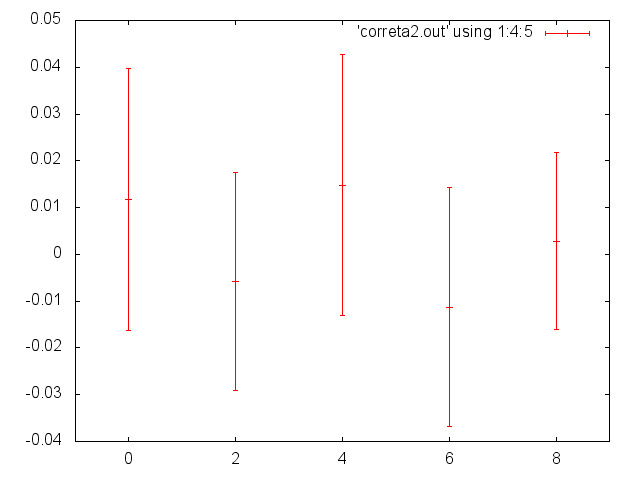}}
\caption[]{Checking the identities, satisfied by the noise fields, as functions of the scalars, for the $\mathcal{N}=2$ two-dimensional Wess-Zumino model, on a 17$\times$ 17 lattice. Left panel: The diagonal components of $\langle\eta_I(x)\eta_J(x')\rangle$ are $\delta-$functions,  up to lattice artifacts; right panel: The off-diagonal components of $\langle\eta_I(x)\eta_J(x')\rangle$ are consistent with 0 across the lattice.}
\label{WZN=2D=2}
\end{figure}

If these identities hold, then the Jacobian--along with its sign--is reproduced by the fluctuations. If they don't, then it isn't. This calculation was done and a sample is shown in fig.~\ref{WZN=2D=2}~\cite{Nicolis:2017lqk}
The fermions are ``hidden'' in the Jacobian, therefore all the correlation functions that involve them can be computed in terms of--very complicated expressions--of the scalar fields. The point being that the ``hard'' part of the calculation is the sampling of the action, which, only, involves the scalars, when done numerically--which is the opposite of what happens when trying to do the corresponding calculations in perturbation theory. 

\section{Beyond two dimensions: doubling the degrees of freedom}\label{beyond2d}
What Parisi and Sourlas remarked was that trying to generalize the approach to spacetime dimensions greater than 2 hit an obstacle: It seemed impossible to write down a superpotential, that had non-quadratic terms and was invariant under SO($d$) rotations, when $d>2,$ in particular when $d=4,$ if one started from the Langevin equation. Stated differently, the Nicolai map didn't seem to exist, for $d>2,$ for Wess-Zumino models.  

What wasn't appreciated was that the real problem, that made, in fact,  the obstruction inevitable, is that, in Euclidian signature, the matrices, $\sigma^\mu,$ that generate the Clifford algebra, have real entries only when $d\equiv\,2\,\mathrm{mod}\,8.$ Therefore the Nicolai map--equivalently the Langevin equation~(\ref{langevin-nicolai})--is incomplete, unless $d\equiv\,2\,\mathrm{mod}\,8$ because the LHS is assumed to be real, while the RHS can't be in $d=4.$

The solution is to double the degrees of freedom: The noise fields are $\eta_I(x)$ and their complex conjugate, $\eta_I^\dagger(x)$ and they are expressed in terms of the $\phi_I(x)$ and their complex conjugate, $\phi_I^\dagger(x),$ according to eqs.~(\ref{langevin-nicolai}) and~(\ref{doublers}). 

Now what is interesting is that this doubling of the degrees of freedom doesn't automatically ensure the invariance of the kinetic term and of the potential term under SO($d$) coordinate transformations and SO($d$) flavor transformations: The non-trivial requirement is that the scalar potential, 
\begin{equation}
\label{VSOd}
V=\frac{1}{2}\frac{\partial W}{\partial\phi_I}\frac{\partial W}{\phi_J^\dagger}\delta^{IJ}
\end{equation}
is a function of $\phi_I\phi_I^\dagger$ (summation over the repeated indices implied);
 indeed this is how the action for $\mathcal{N}=2$ Wess--Zumino models is constructed. This does hold for $D=2.$
 
 Another non--trivial statement is that the expression
\begin{equation}
\label{totderWZ}
\begin{array}{l}
\displaystyle
\mathcal{C}\equiv 
\sigma_{IK}^\mu\partial_\mu\phi_K\left(\frac{\partial W}{\partial\phi_I}\right)^\dagger + \sigma_{LI}^\nu\partial_\nu\phi_L^\dagger\frac{\partial W}{\partial\phi_I}
\end{array}
\end{equation}
that does appear upon expanding the action and isn't, manifestly, invariant under SO($d$) coordinate transformations or under SO($d$) flavor transformations, is, indeed,  a total derivative. It is, obviously, a total derivative if $W$ is a quadratic function of the fields; what remains to check is that non-quadratic terms in the superpotential contribute total derivatives to $\mathcal{C},$ also.

What the doubling of the degrees of freedom achieves is providing the additional terms that are necessary for completing the construction of   the total derivatives. What isn't, yet, proved is, whether this is, also, sufficient. 

It, also, implies that the number of ``flavors'' can't be less than two (four real scalars) in three dimensions and four (eight real scalars) in four dimensions--and these numbers are, simply, the expression of the consistent closure of the system, in equilibrium with its fluctuations, that are resolved by the superpartners. 

\section{Gauge theories}\label{gauge}
Gauge theories probe the group manifold of the corresponding Lie group. For compact gauge groups--that are of relevance to particle physics, in the absence of gravity--the ``natural'' measure is the uniform distribution on the group manifold, that's normalizable. This would be the analog of the Gaussian distribution, with ultra--local 2-point function, on a non-compact manifold, that was used for the Wess--Zumino model. This approach has, in fact, been implemented in recent years, in the framework of lattice gauge theories, more specifically for lattice QCD, using the notion of the  ``trivializing map''. For non-Abelian gauge groups the construction of this map is quite involved and, still, work in progress.  

For Abelian gauge groups, in particular the U(1) group, it is possible to realize the construction of the ``trivializing map''  in the following way, that renders the relation with the Nicolai map more direct. The starting point is the remark that the action of $D$ massless scalar fields, in $D$ dimensions, in Euclidian signature,
\begin{equation}
\label{Smasslessphi}
S[\phi_I]=\int\,d^Dx\,\left\{-\frac{1}{2}\phi_I\Box\phi_J\delta^{IJ}\right\}
\end{equation}
is identical to that of a U(1) gauge field, in $D$ dimensions (in Euclidian signature), in Lorenz-Feynman gauge; just the names have changed: $\phi_I\equiv A_\mu$ with 
$I=1,2,\ldots,D$ and $\mu=1,2,\ldots,D.$ 

This, immediately, implies that the Nicolai map is given by the expression
\begin{equation}
\label{nicolaiU1}
\eta_I =\sigma_{IJ}^\mu\partial_\mu\phi_J
\end{equation}
What is interesting to note is that, if $D\equiv\hskip-0.29truecm/\,2\,\mathrm{mod}\,8,$ we are ``forced'' to double the degrees of freedom, by introducing 
\begin{equation}
\label{nicolaiU1dual}
\eta_I^\dagger=\sigma_{JI}^\mu\partial_\mu\phi_J^\dagger
\end{equation} 
which leads to a natural way of introducing the dual gauge potential. 

The upshot of this analysis is the construction of the Nicolai map for a U(1) gauge theory in $8k+2$ Euclidian dimensions, where the generators of the Clifford algebra admit a Majorana representation and the identification of the photino, upon introducing $\sigma^\mu\partial_\mu$ in the action:
\begin{equation}
\label{photino}
\begin{array}{l}
\displaystyle
Z=\int\,[\mathscr{D}\eta_I]\,e^{-\int\,d^Dx\,\frac{1}{2}\eta_I\eta_J\delta^{IJ}}=1=
\int\,[\mathscr{D}\phi_J]\,\left|\mathrm{det}\,\frac{\delta\eta_I}{\delta\phi_J}\right|\,e^{-\int\,d^Dx\,\left\{-\frac{1}{2}\phi_I\Box\phi_J\delta^{IJ}\right\}}=\\
\displaystyle
\int\,[\mathscr{D}\phi_J][\mathscr{D}\psi_I][\mathscr{D}\chi_J]\,e^{-\mathrm{i}\theta_\mathrm{det}}\,
e^{-\int\,d^Dx\,\left\{-\frac{1}{2}\phi_I\Box\phi_J\delta^{IJ}-\psi_I\sigma_{IJ}^\mu\partial_\mu\chi_J\right\}}
\end{array}
\end{equation} 
We note that the phase of the Jacobian, $e^{-\mathrm{i}\theta_\mathrm{det}},$ is independent of the fields (for abelian gauge fields, in the absence of matter and of boundaries) and can, thus, be taken out of the integral. It's a global phase in this case. 

We remark the doubling of anticommuting degrees of freedom, in order to represent the determinant of $\sigma^\mu\partial_\mu$ as a ``local'' term in the Euclidian action. 

It is interesting to remark now that we can interpret the $I,J$ indices as, either, ``flavor'' indices, or as ``spacetime'' indices.

In the first case, we have a theory of $D$ scalar multiplets, that describe $D$ scalar fields, while their superpartners describe the fluctuations. 

In the second case, we have a theory of an abelian gauge field in $D$ dimensions and its superpartners, the gaugini, that describe the fluctuations. 

Which case is relevant depends, of course, on additional constraints,  that are introduced (indeed this is one way to describe fields of any spin).

At this point it's useful to remark that the term $\sigma_{IJ}^\mu\partial_\mu\phi_J,$ displays one way of coupling $D$ scalar multiplets in a way consistent with invariance under SO($D$) transformations and supersymmetry. 

Therefore we have a consistent description of the photon and the photino, in $D=8k+2$ dimensions. However these are free fields, so the question is, how could they interact. The answer is, precisely, by making explicit the property that they are superpartners: By measuring properties of the photon, it is possible to deduce, first of all, the inevitable existence of the photino and, second, its properties.

However, precisely because these are free fields, their properties can only be probed upon coupling them to other fields. Such fields are matter fields, charged under the gauge field. The obvious {\em Ansatz} for the matter fields is in terms of the Langevin equations,
\begin{equation}
\label{Langevinmatt}
\begin{array}{l}
\displaystyle
\xi_I=\sigma_{IJ}^\mu\nabla_\mu\varphi_J+\frac{\partial W}{\partial\varphi_I}\\
\displaystyle
\xi_I^\dagger =\sigma_{JI}^\mu[\nabla_\mu\varphi_J]^\dagger+\left(\frac{\partial W}{\partial\varphi_I}\right)^\dagger
\end{array}
\end{equation}
where 
\begin{equation}
\label{covder}
\nabla_\mu\equiv\partial_\mu-\mathrm{i}q A_\mu\equiv\partial_\mu-\mathrm{i}q\phi_\mu
\end{equation}  
where $\phi_\mu\equiv\phi_I\equiv A_\mu$ and $q$ is the charge of the matter fields under the gauge field.

The partition function is, therefore, defined by 
\begin{equation}
\label{Zsqed}
\begin{array}{l}
\displaystyle
Z=\int\,\underbrace{[\mathscr{D}\eta_I][\mathscr{D}\xi_I][\mathscr{D}\xi_I^\dagger]}_{[\mathscr{D}h_I]}\,
e^{-\int\,d^Dx\,\left\{ \frac{1}{2}\eta_I\eta_J\delta^{IJ}+\xi_I\xi_J^\dagger\delta^{IJ}\right\}}=1=\\
\displaystyle
\int\,\underbrace{[\mathscr{D}\phi_I][\mathscr{D}\varphi_I][\mathscr{D}\varphi_I^\dagger]}_{[\mathscr{D}\Phi_I]}\,
\left|\mathrm{det}\,\frac{\delta h_I}{\delta\Phi_J}\right|\,e^{-S[\phi_I,\varphi_I,\varphi_I^\dagger]}
\end{array}
\end{equation}
If we expand out the action, replace $\phi_I\to A_\mu$ in order to conform to standard notation and introduce the determinant in the action using anticommuting fields it is clear that we will obtain the partition function of  supersymmetric QED--with extended supersymmetry--in $d=8k+2$ spacetime dimensions. In fact the presence of the phase of the determinant implies that this expression can be identified with the Witten index. This provides a new way of coupling U(1) gauge theory to matter within the framework of extended supersymmetry, beyond what  was studied in refs.~\cite{Wess:1974jb,Elitzur:1982vh,Ambjorn:1997cq,Ambjorn:1997yw}. 

Including more  ``flavors'' is, of course, obvious. What is quite interesting, once more, is that this approach implies the {\em least} number of flavors, that are required for the system to be consistently closed and that this number, for worldvolumes of more than one dimensions and, in order to describe target space fermions, is greater than 1 and, indeed, prescribed by the requirement that the Langevin equation define an action that's bounded from below and confines at infinity.

It must be stressed that this expression for the partition function is exact and describes {\em all} the fluctuations of the fields that appear in the classical action, by virtue of the fact that it is equal to 1. What this implies, in particular, is that, if the classical action is taken to be the action of the scalars, coupled to the gauge field, then the fluctuations 
are described by the fermions--along with the phase of the determinant. Conversely, if the classical action is taken to be the action of the fermions, coupled to the gauge field, then the fluctuations are described by the scalars--along with the phase of the determinant. In both cases, the fluctuations of the gauge field are described by the part of the 
determinant, $\mathrm{det}\,(\delta h_I/\delta\Phi_J),$ that won't, however, factorize, in general, in a ``local'' way. 

However it's not necessary to deal with the absolute value of the determinant directly--that's where the Nicolai maps enter the picture, as well as the knowledge that this is, in fact,  a supersymmetric theory--it is supersymmetry that relates the correlation functions of the scalar fields, that are easy to compute, to the correlation functions of the fermions, that are much harder to, directly, compute.

The Nicolai maps~(\ref{nicolaiU1}),(\ref{nicolaiU1dual}) and~(\ref{Langevinmatt}) highlight the interesting observables of the theory; they indicate that the corresponding fields, when sampled using the measure of the ``canonically'' defined gauge theory, in Lorenz--Feynman gauge, should be Gaussian, with ultra--local 2--point function; if the 1--point functions vanish, supersymmetry is intact, otherwise it's spontaneously broken. If the 2--point function isn't ultra--local, or the $\eta_I,\xi_I,\xi_I^\dagger$ aren't Gaussian, supersymmetry is ``anomalously'' broken--and the deviation of the identities satisfied by these ``noise fields'' from what's expected from Wick's theorem, allows a quantitative description of this breaking.  In four spacetime dimensions we note that perturbative renormalizability imposes that the superpotential should be at most cubic, which leads, inevitably, to supersymmetry being, at least, spontaneously,  broken.

It should be noted that, in spacetime dimensions other than $8k+2$ (in particular in the physically interesting cases of $D=3$ and $D=4$), it's necessary to double the degrees of freedom--so the dual fields, that  can realize electric/magnetic duality~\cite{Seiberg:1994rs,Bilal:1995hc} appear naturally. 

Furthermore, we remark that this formulation shows how to describe the quantum effects of theories of non-chiral fermions, charged under an abelian U(1) gauge field, namely by $\mathcal{N}=2$ SQED: the additional scalars and the photino provide the consistent closure, since including them leads to the identification of the partition function with the Witten index. This provides the framework for experiments that can probe the effects of the superpartners of the electron--the selectron--and the photon--the photino--in traps and in current magnetic materials, where fluctuations play an increasingly  important role.

So it remains to provide the  numerical evidence that the correlation functions for the noise fields, $\eta_I$ and $\xi_I,\xi_I^\dagger,$ when expressed in terms of the scalars $\phi_I\equiv A_\mu,$ $\varphi_I,\varphi_I^\dagger,$ do indeed satisfy  the identities expected of Gaussian fields, with ultra--local 2--point function. The most surprising identity is that which shows factorization, namely, $\langle\eta_I\xi_J\rangle=0=\langle\eta_I\xi_J^\dagger\rangle.$ Of course neither the $\eta_I$ or the $\xi_I,\xi_I^\dagger$ are gauge invariant quantities; nevetheless the factorization property is a gauge invariant statement, since it expresses charge conservation.

Let us, therefore, close by addressing the topic of investigations using numerical simulations and a lattice regularization. The Nicolai maps imply that it suffices to work with the lattice action for the scalars and the gauge fields; the effects of supersymmetry are described by the identities of the correlation functions of the noise fields. So, for the case at hand, it suffices to work with the action of the compact Abelian--Higgs model (with the requisite number of scalars). This has the advantage of eliminating any conceptual issues and reducing the problem to a case where known and tested numerical techniques can be brought to bear. However we realize that the gauge fields themselves can be described in terms of scalars and it is the correlation functions that reveal which scalars are, in fact, gauge fields.

The action for the fields $\phi_I$ and $\varphi_I,\varphi_I^\dagger$ is readily obtained, by replacing the expressions for the noise fields in the action. We thus obtain the expression
\begin{equation}
\label{SclassSQED}
S[\phi_I,\varphi_I,\varphi_I^\dagger]=\int\,d^Dx\,\left\{
-\frac{1}{2}\phi_I\Box\phi_J\delta^{IJ}+\left[\sigma_{IK}^\mu\nabla_\mu\varphi_K+\frac{\partial W}{\partial\varphi_I}\right]
\left[\sigma_{JL}^\mu[\nabla_\mu\varphi_L]^\dagger+\left(\frac{\partial W}{\partial\varphi_J}\right)^\dagger\right]\delta^{IJ}
\right\}
\end{equation}
 it is bounded from below and confines at infinity  and its lattice discretization is straightforward. It seems to define the dynamics of scalar fields only; the non-trivial properties are encoded by the correlation functions of the noise fields. 

The simplest case is that of $D=2$ Euclidian spacetime dimensions. The Langevin equations presented above realize the  coupling of the $\mathcal{N}=2$ Wess-Zumino model to photon and photino. 
That we're studying electrodynamics is ``hidden'' in the fact that we're calculating correlation functions of ``gauge-invariant'' quantities, namely functions of $F_{IJ}=\partial_I\phi_J-\partial_J\phi_I$ and of ``Wilson lines'' that end on $\varphi_I$ fields.   

(The noise fields are not invariant under gauge transformations--it's their bilinears that are.)

Of course the outstanding question is the generalization to non-abelian gauge fields. The Nicolai map for the matter fields (that generalize eqs.~(\ref{Langevinmatt}) ) can be immediately written, whether in the continuum or on the lattice, since the expression for the covariant derivative (in the continuum) or the ``covariant link'' (on the lattice) is known; what isn't, for the moment, clear is what the Nicolai map for the gauge fields should be, in Lorenz--Feynman gauge (or any gauge, in fact).  If we attempt to generalize the approach used for abelian gauge fields, we realize that where it will fail is in the definition of the potential term, since the term that describes the self--coupling of non-abelian gauge fields, the  commutator, $[A_\mu^a,A_\nu^b],$ can't be written as a gradient.  There are, still, conceptual issues that need to be clarified for the construction of non-abelian gauge theories in this way to be effectively realized.

Let us, indeed, point out that we worked with fields taking values in the Lie algebra of the gauge group. The gauge fields, that describe the non-gravitational interactions of matter, whether within the framework of the Standard Model or in its generalizations, take values in the Lie algebra of a compact group.  While the algebra, describing the vicinity of the identity, is probed within perturbation theory about free fields, 
the full group is, of course, of physical interest and is accessible by working on the lattice, where the gauge fields naturally take values in the group. It is well known that current algorithms that attempt to sample the group manifold have limitations (in particular critical slowing down) upon attempting to sample different topological sectors. This has led to 
the idea of ``trivializing maps'', that transform any measure $e^{-S[U]}[dU]$ on the group manifold to the uniform measure, $[dV],$ on the same. There is, of course, a corresponding Jacobian, defined by  $e^{-S[U]} dU|dV/dU|=dV.$ The construction of this Jacobian was presented in ref.~\cite{luscher2010trivializing} and a non-trivial difference with the scalar case is that its construction is considerably more involved than that of the Nicolai map.   The exponentiation of the Jacobian, however,  leads to the appearance of the gaugini in the action where they are expected to do so. The uniform measure on compact manifolds plays the same role as that of  the  Gaussian measure with 
$\delta-$function variance on non-compact manifolds. What is remarkable is that, for abelian gauge fields, it is possible to avoid the complications and realize that these reflect the particular properties of the self-interactions of  non-abelian gauge fields, when they take values in the algebra. It will be interesting to understand the relation with the recent work on the Nicolai map presented in~\cite{Lechtenfeld:2022qpa,Catterall:2022qzs,Lechtenfeld:2021uvs,Ananth:2020gkt}.

\section{Conclusions}\label{concl}
For many years, since the invention of supersymmetry, it has been considered an option for describing physical systems. While it does lead to many simplifications for technical calculations,  there hadn't seemed  to be any particular physical reason for its presence  to be considered inevitable. The simplifications didn't seem to imply that, in their absence, the mathematical formulation of physical phenomena was inconsistent--it was more cumbersome and, at best, incomplete; but it did seem  possible to describe ``non-supersymmetric'' systems consistently. The presence of the superpartners of the known particles didn't appear to be mandatory. If they were indeed present, they could play a very useful role; but there didn't seem to be any compelling reason for them to be present. The discovery of the Higgs boson does imply that new particles must exist-but it doesn't imply, for the moment, anything about how they may appear in supermultiplets. 

 It was only in 1982 that Parisi and Sourlas~\cite{parisi_sourlas} proposed that supersymmetry describes a fundamental physical property of {\em any} physical system, namely that it was consistently closed, when in equilibrium with a bath and, more precisely,  they provided strong arguments that  the superpartners of the one-particle excitations of scalar fields could describe the one-particle excitations that could resolve the fluctuations of the bath, with which the fields, that appeared in the classical action, are in equilibrium. 
  They thus provided a compelling reason for taking into account supersymmetry as an inevitable property, rather than an option, when describing a system in equilibrium with a bath.  More strikingly, the essence of the proposal was that it was supersymmetry that expressed the property that how the system was separated into ``dynamical'' degrees of freedom and degrees of freedom of the ``bath of fluctuations'', didn't matter and shouldn't matter. If it did matter, supersymmetry was broken--but that implied that  the description of the system was incomplete. 
 
However, while  they provided  the framework for understanding that the superpartners can describe the fluctuations, one question, that was left unanswered in their paper, was, whether  the fluctuations of the dynamical degrees of freedom could, indeed, produce the effects of their superpartners, or whether the superpartners had to be introduced separately. In other words, are all theories, actually, supersymmetric, they just don't seem to be, due to the approximation of focusing on certain degrees of freedom and ignoring the others,  or are only supersymmetric theories closed in this sense? 

Addressing these issues requires  going beyond perturbation theory, therefore performing numerical simulations. However a prerequisite is identifying what would be the quantities, whose calculation would be relevant for addressing--and answering--these questions. 

In fact these had been identified by Nicolai~\cite{Nicolai:1980jc,Nicolai:1980js}, who proposed what became known subsequently as the ``Nicolai map''. The idea was that, in  a supersymmetric theory, it is possible to describe the anticommuting superpartners of the commuting fields of a supermultiplet as functions of the commuting fields. The correspondence isn't direct: The anticommuting partners ``emerge'' upon introducing into the action for the commuting fields, a certain determinant, that is a function of the commuting fields. 
This map was found for the case of Wess--Zumino models. What wasn't realized is that this map is nothing more or less than the Langevin equation, that was the starting point of the proposal of Parisi and Sourlas. However, while there's been a lot of work on the two-dimensional Wess-Zumino model, the three-- and four--dimensional models have received much less attention and the Nicolai map hasn't been studied as fully. One reason may have to do with the obstruction found by Parisi and Sourlas. The way around it (presented in ref.~\cite{Nicolis:2021buh}) provides a necessary, but not sufficient, condition for the definition of the Nicolai map and more work is needed to spell out the details.   

The determinant indeed, represents a change of variables in the path integral, from the dynamical fields to ``noise fields'', whose correlation functions define the bath--so its sign (more generally its phase) must be taken into account as well. With hindsight it is possible to identify the quantity, that contains the phase of the determinant with the so--called  Witten index~\cite{Witten:1981nf,Witten:1982im,Witten:1982df,Witten:2015aba} and realize that  the whole point of the proposal was that, when the fluctuations are fully taken into account, the canonical partition function becomes, in fact, the Witten index.  

In this contribution we have provided numerical evidence that, for the $\mathcal{N}=2,D=2$ Wess-Zumino model, the proposal by Parisi and Sourlas does work: The fluctuations of the scalars can be repackaged as the absolute value of the determinant, that realizes the change of variables to the noise fields. Further simulations that support this picture have been carried out in refs.~\cite{Morikawa:2018ops,Morikawa:2018zys,Morikawa:2019cql} that, also, probe the relation with Landau-Ginzburg models. 
Furthermore, we have spelled out the framework for realizing the proposal for abelian gauge theories coupled to matter fields, thereby leading to the description of $\mathcal{N}=2$ SQED. In what way supersymmetry can be realized or broken can thus be addressed by numerical simulations of the corresponding scalar theory, with the Nicolai maps providing the appropriate observables. While the simulations are challenging, they are well-defined and will be reported in future work. 

Another challenge pertains to describing chiral fermions in this approach, since $\mathcal{N}=2$ theories describe non-chiral fermions. One way might be  using partial breaking of $\mathcal{N}=2\to \mathcal{N}=1$~\cite{Antoniadis:1995vb}. It would be interesting to see whether $\mathcal{N}=1$ models that were studied in ref.~\cite{Heinemeyer:2021zrk} could be obtained this way. 
 
{\bf Acknowledgements:} It's a pleasure to thank the organizers of the Workshops held at the Corfu Summer Institute for organizing a wonderful conference. Discussions with  M. Axenides, E. Floratos and J. Iliopoulos are gratefully acknowledged. 
\bibliographystyle{JHEP}
\bibliography{SUSY}

\providecommand{\href}[2]{#2}\begingroup\raggedright\begin{thebibliography}{10}

\bibitem{parisi_sourlas}
G.~Parisi and N.~Sourlas, \emph{{Supersymmetric Field Theories and Stochastic
  Differential Equations}},
  \href{http://dx.doi.org/10.1016/0550-3213(82)90538-7}{\emph{Nucl. Phys.}
  {\bfseries B206} (1982) 321--332}.

\bibitem{Nicolis:2017lqk}
S.~Nicolis, \emph{{Probing the holomorphic anomaly of the $D=2, \mathcal{N}=2$,
  Wess-Zumino model on the lattice}},
  \href{http://dx.doi.org/10.1134/S1063779618050313}{\emph{Phys. Part. Nucl.}
  {\bfseries 49} (2018) 899--903},
  [\href{https://arxiv.org/abs/1712.07045}{{\ttfamily 1712.07045}}].

\bibitem{Wess:1974jb}
J.~Wess and B.~Zumino, \emph{{Supergauge Invariant Extension of Quantum
  Electrodynamics}},
  \href{http://dx.doi.org/10.1016/0550-3213(74)90112-6}{\emph{Nucl. Phys. B}
  {\bfseries 78} (1974) 1}.

\bibitem{Elitzur:1982vh}
S.~Elitzur, E.~Rabinovici and A.~Schwimmer, \emph{{Supersymmetric Models on the
  Lattice}}, \href{http://dx.doi.org/10.1016/0370-2693(82)90269-6}{\emph{Phys.
  Lett. B} {\bfseries 119} (1982) 165}.

\bibitem{Ambjorn:1997cq}
J.~Ambjorn, N.~Sasakura and D.~Espriu, \emph{{U(1) lattice gauge theory and
  $\mathcal{N}=2$ supersymmetric Yang-Mills theory}}, {\emph{Fortsch. Phys.}
  {\bfseries 47} (1999) 287--292}.

\bibitem{Ambjorn:1997yw}
J.~Ambjorn, D.~Espriu and N.~Sasakura, \emph{{U(1) lattice gauge theory and
  $\mathcal{N}=2$ supersymmetric Yang-Mills theory}},
  \href{http://dx.doi.org/10.1142/S0217732397002806}{\emph{Mod. Phys. Lett. A}
  {\bfseries 12} (1997) 2665--2682},
  [\href{https://arxiv.org/abs/hep-th/9707095}{{\ttfamily hep-th/9707095}}].

\bibitem{Seiberg:1994rs}
N.~Seiberg and E.~Witten, \emph{{Electric - magnetic duality, monopole
  condensation, and confinement in N=2 supersymmetric Yang-Mills theory}},
  \href{http://dx.doi.org/10.1016/0550-3213(94)90124-4}{\emph{Nucl. Phys. B}
  {\bfseries 426} (1994) 19--52},
  [\href{https://arxiv.org/abs/hep-th/9407087}{{\ttfamily hep-th/9407087}}].

\bibitem{Bilal:1995hc}
A.~Bilal, \emph{{Duality in $\mathcal{N}=2$ SUSY SU(2) Yang-Mills theory: A
  Pedagogical introduction to the work of Seiberg and Witten}},  in \emph{{NATO
  Advanced Study Institute on Quantum Fields and Quantum Space Time}},
  pp.~21--43, 1997.
\newblock \href{https://arxiv.org/abs/hep-th/9601007}{{\ttfamily
  hep-th/9601007}}.

\bibitem{luscher2010trivializing}
M.~L{\"u}scher, \emph{{Trivializing maps, the Wilson flow and the HMC
  algorithm}}, {\emph{Communications in Mathematical Physics} {\bfseries 293}
  (2010) 899}.

\bibitem{Lechtenfeld:2022qpa}
O.~Lechtenfeld and M.~Rupprecht, \emph{{Is the Nicolai map unique?}},
  \href{https://arxiv.org/abs/2207.09471}{{\ttfamily 2207.09471}}.

\bibitem{Catterall:2022qzs}
S.~Catterall and J.~Giedt, \emph{{Supersymmetric Lattice Theories: Contribution
  to Snowmass 2022}},  2, 2022.
\newblock \href{https://arxiv.org/abs/2202.08154}{{\ttfamily 2202.08154}}.

\bibitem{Lechtenfeld:2021uvs}
O.~Lechtenfeld and M.~Rupprecht, \emph{{Universal form of the Nicolai map}},
  \href{https://arxiv.org/abs/2104.00012}{{\ttfamily 2104.00012}}.

\bibitem{Ananth:2020gkt}
S.~Ananth, H.~Nicolai, C.~Pandey and S.~Pant, \emph{{Supersymmetric
  Yang\textendash{}Mills theories:not quite the usual perspective}},
  \href{http://dx.doi.org/10.1088/1751-8121/ab7b9d}{\emph{J. Phys. A}
  {\bfseries 53} (2020) 17},
  [\href{https://arxiv.org/abs/2001.02768}{{\ttfamily 2001.02768}}].

\bibitem{Nicolai:1980jc}
H.~Nicolai, \emph{{Supersymmetry without anticommuting variables}},  in
  \emph{{Unification of the fundamental particle interactions. Proceedings,
  Europhysics study conference, Erice, Italy, March 17-24, 1980}}, p.~689,
  1980.

\bibitem{Nicolai:1980js}
H.~Nicolai, \emph{{Supersymmetry and Functional Integration Measures}},
  \href{http://dx.doi.org/10.1016/0550-3213(80)90460-5}{\emph{Nucl. Phys.}
  {\bfseries B176} (1980) 419--428}.

\bibitem{Nicolis:2021buh}
S.~Nicolis, \emph{{The hidden fluxes, that control the fluctuations of scalar
  fields}}, \href{http://dx.doi.org/10.1088/1742-6596/2105/1/012003}{\emph{J.
  Phys. Conf. Ser.} {\bfseries 2105} (2021) 012003},
  [\href{https://arxiv.org/abs/2107.03194}{{\ttfamily 2107.03194}}].

\bibitem{Witten:1981nf}
E.~Witten, \emph{{Dynamical Breaking of Supersymmetry}},
  \href{http://dx.doi.org/10.1016/0550-3213(81)90006-7}{\emph{Nucl. Phys.}
  {\bfseries B188} (1981) 513}.

\bibitem{Witten:1982im}
E.~Witten, \emph{{Supersymmetry and Morse theory}}, {\emph{J. Diff. Geom.}
  {\bfseries 17} (1982) 661--692}.

\bibitem{Witten:1982df}
E.~Witten, \emph{{Constraints on Supersymmetry Breaking}},
  \href{http://dx.doi.org/10.1016/0550-3213(82)90071-2}{\emph{Nucl. Phys.}
  {\bfseries B202} (1982) 253}.

\bibitem{Witten:2015aba}
E.~Witten, \emph{{Fermion Path Integrals And Topological Phases}},
  \href{http://dx.doi.org/10.1103/RevModPhys.88.035001,
  10.1103/RevModPhys.88.35001}{\emph{Rev. Mod. Phys.} {\bfseries 88} (2016)
  035001}, [\href{https://arxiv.org/abs/1508.04715}{{\ttfamily 1508.04715}}].

\bibitem{Morikawa:2018ops}
O.~Morikawa and H.~Suzuki, \emph{{Numerical study of the $\mathcal{N}=2$
  Landau–Ginzburg model}},
  \href{http://dx.doi.org/10.1093/ptep/pty088}{\emph{PTEP} {\bfseries 2018}
  (2018) 083B05}, [\href{https://arxiv.org/abs/1805.10735}{{\ttfamily
  1805.10735}}].

\bibitem{Morikawa:2018zys}
O.~Morikawa, \emph{{Numerical study of the $\mathcal{N}=2$ Landau--Ginzburg
  model with two superfields}},
  \href{http://dx.doi.org/10.1007/JHEP12(2018)045}{\emph{JHEP} {\bfseries 12}
  (2018) 045}, [\href{https://arxiv.org/abs/1810.02519}{{\ttfamily
  1810.02519}}].

\bibitem{Morikawa:2019cql}
O.~Morikawa, \emph{{Continuum limit in numerical simulations of the
  $\mathcal{N}=2$ Landau--Ginzburg model}},
  \href{https://arxiv.org/abs/1906.00653}{{\ttfamily 1906.00653}}.

\bibitem{Antoniadis:1995vb}
I.~Antoniadis, H.~Partouche and T.~R. Taylor, \emph{{Spontaneous breaking of
  N=2 global supersymmetry}},
  \href{http://dx.doi.org/10.1016/0370-2693(96)00028-7}{\emph{Phys. Lett. B}
  {\bfseries 372} (1996) 83--87},
  [\href{https://arxiv.org/abs/hep-th/9512006}{{\ttfamily hep-th/9512006}}].

\bibitem{Heinemeyer:2021zrk}
S.~Heinemeyer, J.~Kalinowski, W.~Kotlarski, M.~Mondragon, G.~Patellis,
  N.~Tracas et~al., \emph{{Probing a Finite Unified Theory with Reduced
  Couplings at Future Colliders}},
  \href{http://dx.doi.org/10.22323/1.398.0736}{\emph{PoS} {\bfseries
  EPS-HEP2021} (2022) 736}, [\href{https://arxiv.org/abs/2110.07261}{{\ttfamily
  2110.07261}}].

\end{thebibliography}\endgroup


\end{document}